\documentclass[twocolumn]{aastex631}

\newcommand{\Mso}{$M_{\odot}$}

\usepackage{graphicx}	% Including figure files
\usepackage{amsmath}	% Advanced maths commands
\usepackage{amssymb}	% Extra maths symbols
\usepackage{enumitem}
\usepackage{hyperref}
\hypersetup{colorlinks=true,allcolors=blue,pdfborder={0 0 0}}
\shorttitle{Helix Nebula}
\shortauthors{Marshall et al.}
\graphicspath{{./}{figures/}}
\begin{document}

\title{Evidence for the disruption of a planetary system during the formation of the Helix Nebula}

\correspondingauthor{Jonathan P. Marshall}
\email{jmarshall@asiaa.sinica.edu.tw}

\author[0000-0001-6208-1801]{Jonathan P. Marshall}
\affiliation{Academia Sinica, Institute of Astronomy and Astrophysics,\\ 11F Astronomy-Mathematics Building, NTU/AS campus, No. 1, Section 4, Roosevelt Rd., Taipei 10617, Taiwan}
\affiliation{Centre for Astrophysics, University of Southern Queensland, Toowoomba, QLD 4350, Australia}

\author[0000-0002-2314-7289]{Steve Ertel}
\affiliation{Large Binocular Telescope Observatory, University of Arizona, 933 N. Cherry Avenue, Tucson, AZ 85721-0065, USA}
\affiliation{Department of Astronomy and Steward Observatory, University of Arizona, 933 N. Cherry Avenue, Tucson, AZ 85721-0065, USA}

\author{Eric Birtcil}
\affiliation{Department of Astronomy and Steward Observatory, University of Arizona, 933 N. Cherry Avenue, Tucson, AZ 85721-0065, USA}

\author[0000-0003-4936-9418]{Eva Villaver}
\affiliation{Centro de Astrobiolog\'ia (CAB, CSIC-INTA), ESAC Campus Camino Bajo del Castillo, s/n, Villanueva de la Ca\~nada, 28692, Madrid, Spain}

\author[0000-0003-2743-8240]{Francisca Kemper}
\affiliation{Institut de Ciencies de l'Espai (ICE, CSIC), Can Magrans, s/n, 08193 Bellaterra, Barcelona, Spain}
\affiliation{ICREA, Pg. Llu\'is Companys 23, Barcelona, Spain}
\affiliation{Institut d'Estudis Espacials de Catalunya (IEEC), E-08034 Barcelona, Spain}

\author[0000-0002-9486-4840]{Henri Boffin}
\affiliation{European Southern Observatory, Alonso de Cordova 3107, Santiago RM, Chile}

\author[0000-0002-1161-3756]{Peter Scicluna}
\affiliation{European Southern Observatory, Alonso de Cordova 3107, Santiago RM, Chile}

\author[0000-0001-8299-3402]{Devika Kamath}
\affiliation{School of Mathematical and Physical Sciences, Macquarie University, Sydney, NSW 2118, Australia}
\affiliation{Astronomy, Astrophysics and Astrophotonics Research Centre, Macquarie University, Sydney, NSW 2118, Australia}



\begin{abstract}

The persistence of planetary systems after their host stars evolve into their post-main sequence phase is poorly constrained by observations. Many young white dwarf systems exhibit infrared excess emission and/or spectral absorption lines associated with a reservoir of dust (or planetesimals) and its accretion. However, most white dwarfs are too cool to sufficiently heat any circumstellar dust to detectable levels of emission. The Helix Nebula (NGC 7293) is a young, nearby planetary nebula; observations at mid- and far-infrared wavelengths revealed excess emission associated with its central white dwarf (WD 2226-210). The origin of this excess is ambiguous. It could be a remnant planetesimal belt, a cloud of comets, or the remnants of material shed during the post-asymptotic giant branch phase. Here we combine infrared (SOFIA, \textit{Spitzer}, \textit{Herschel}) and millimetre (ALMA) observations of the system to determine the origin of this excess using multi-wavelength imaging and radiative transfer modelling. We find the data are incompatible with a compact remnant planetesimal belt or post-asymptotic giant branch disc, and conclude the dust most likely originates from deposition by a cometary cloud. The measured dust mass, and lifetime of the constituent grains, implies disruption of several thousand Hale-Bopp equivalent comets per year to fuel the observed excess emission around the Helix Nebula's white dwarf.

\end{abstract}



\section{Introduction} 
\label{sec:intr}

Our understanding of the afterlives of planetary systems, once the host star has evolved off the main sequence, is predominantly based on modelling \citep[e.g.][]{2009Villaver, 2012Mustill,2017Veras,2019Veras, 2018Mustill,2014Villaver,2020Veras,2020Maldonado}. Observing the architectures of planetary systems around white dwarfs would critically constrain these models. We have various lines of indirect evidence that such systems exist, including infrared excess emission \citep[e.g.][]{2007Jura,2012Xu,2015Xu} and contamination of white dwarf spectra with metal absorption lines \citep[e.g.][]{2006Jura,2013Xu,2014Xu}. \textit{Spitzer} mid-infrared surveys of hot white dwarfs found $\simeq$~20~\% of the objects located in planetary nebula exhibit infrared excess \citep{2012Bilikova}.

White dwarf absorption spectra are one of the few ways in which the composition of planetesimals around stars can be determined, yielding important clues to the composition of planetary companions to other stars \citep{2014JuraYoung}. Surviving multiple planets bound to the stellar host are believed to play an important role in scattering planetesimals into the white dwarf's Roche limit producing the observed excesses and contamination \citep[e.g.][]{2002Debes,2021Maldonado}. The recent discovery of Jovian-mass exoplanets around white dwarfs has strengthened the case that the post-main sequence survival of planetary system is indeed possible, and the mechanisms proposed to place material close to the white dwarf are plausible \citep{2019Gansicke,2020Vanderburg,2021Blackman,2022Scaringi}. 

The Helix Nebula (NGC~7293) is a polypolar, rather than bipolar, planetary nebula viewed at a pole-on orientation \citep{2004Odell}.  {Polypolar planetary nebulae exhibit structure associated with multiple, episodic bipolar outflow events at different orientations due to precession of the bipolar outflow \citep[e.g.][]{1974KalerAller,1996Manchado,1998Lopez,2019Hsia}, around 20 per cent of young planetary nebulae are found to be polypolar \citep{2011Sahai}.} Lying at a distance of 201~$\pm$~3~pc, the Helix Nebula is one of the closest such systems to the Sun \citep{2016Gaia,2018Gaia}. Its central white dwarf (WD~2226-210) has a candidate low mass binary companion in a few days orbit based on H$\alpha$ line emission \citep{2001Gruendl} and photometric observations of periodic variability by the \textit{Transiting Exoplanet Survey Satellite} \citep{2020Aller}. The most likely explanation for the variability is found to be irradiation of a substellar or planetary mass body that with a period of a few days that must have gone through common envelope evolution. 

WD~2226-210 is a young, DAO white dwarf with an effective temperature of around $T_{\star} = 103\,600~\pm~5\,500$~K, $\log g = 7.0~\pm~0.2$ \citep{1999Napiwotzki,2005Traulsen}, and an estimated age of 10.6$^{+2.3}_{-1.2}$~kyr based on the expansion velocity of the nebula \citep{2002ODell}.  {More recent results record a slightly lower temperature $T_{\star} = 94\,640~\pm~3\,349~$K and similar surface gravity \citep{2011Gianninas}, but these are consistent with each other. For the purposes of this work we have adopted the values of \cite{1999Napiwotzki} and \cite{2005Traulsen} to model the white dwarf spectrum}. It is a powerful source of X-ray emission, providing a direct constraint on the density of material along the line-of-sight \citep{2015Montez}. 

Binary post-asymptotic giant branch (post-AGB) stars, thought to be the progenitors of bipolar and polypolar planetary nebulae, are surrounded by gas- and dust-rich discs that resemble protoplanetary discs around pre-main sequence stars \citep[e.g., ][]{2006deRuyter,2017Hillen}. These post-AGB discs are sites of dust grain growth \citep{1999Molster,2020Scicluna}, and potentially planet(esimal) formation and growth \citep{2014Volschow,2014BearSoker}. In combination, these properties make the Helix Nebula an excellent candidate to study the post-main sequence fate of circumstellar material.

WD~2226-210 was resolved from the surrounding nebula at infrared wavelengths by \textit{Spitzer}, revealing excess emission consistent with a circumstellar dust disc \citep{2007Su}. Subsequent observations at far infrared wavelengths by \textit{Herschel} supported this finding \citep{2015VandeSteene}, although those observations were focused on the wider nebula and not the white dwarf specifically. The presence of circumstellar emission from the white dwarf points to either the existence of remnant planetesimals from the main sequence progenitor system (either asteroids or comets), a debris disc evolved to the post-main sequence phase \citep{2010Bonsor,2020Veras} or perhaps a secondary disc formed from stellar ejecta during its post-asymptotic giant branch phase if the system is indeed a binary \citep{2022Kluska}.

In this article we focus our efforts on deducing the origin and nature of the unresolved excess emission observed around the central white dwarf of the Helix Nebula. This is part of our wider effort to observationally constrain the post-main sequence evolution of planetary systems \citep[see also][]{2019Ertel,2020Scicluna}. The remainder of the paper proceeds as follows. In Sections \ref{sec:obs} and \ref{sec:meth}, we summarise the observations and describe the methodology used to interpret the multi-wavelength imaging and spectroscopy. Next, in Section \ref{sec:resl}, we present the results of multi-wavelength image analysis and radiative transfer modelling. We then discuss the implications for the origin of the observed excess in Section \ref{sec:disc}, before giving our conclusions in Section \ref{sec:conc}.

\section{Observations}
\label{sec:obs}

\begin{figure*}
    \centering
    \includegraphics[width=0.45\textwidth,trim={0cm 3cm 0cm 0cm},clip]{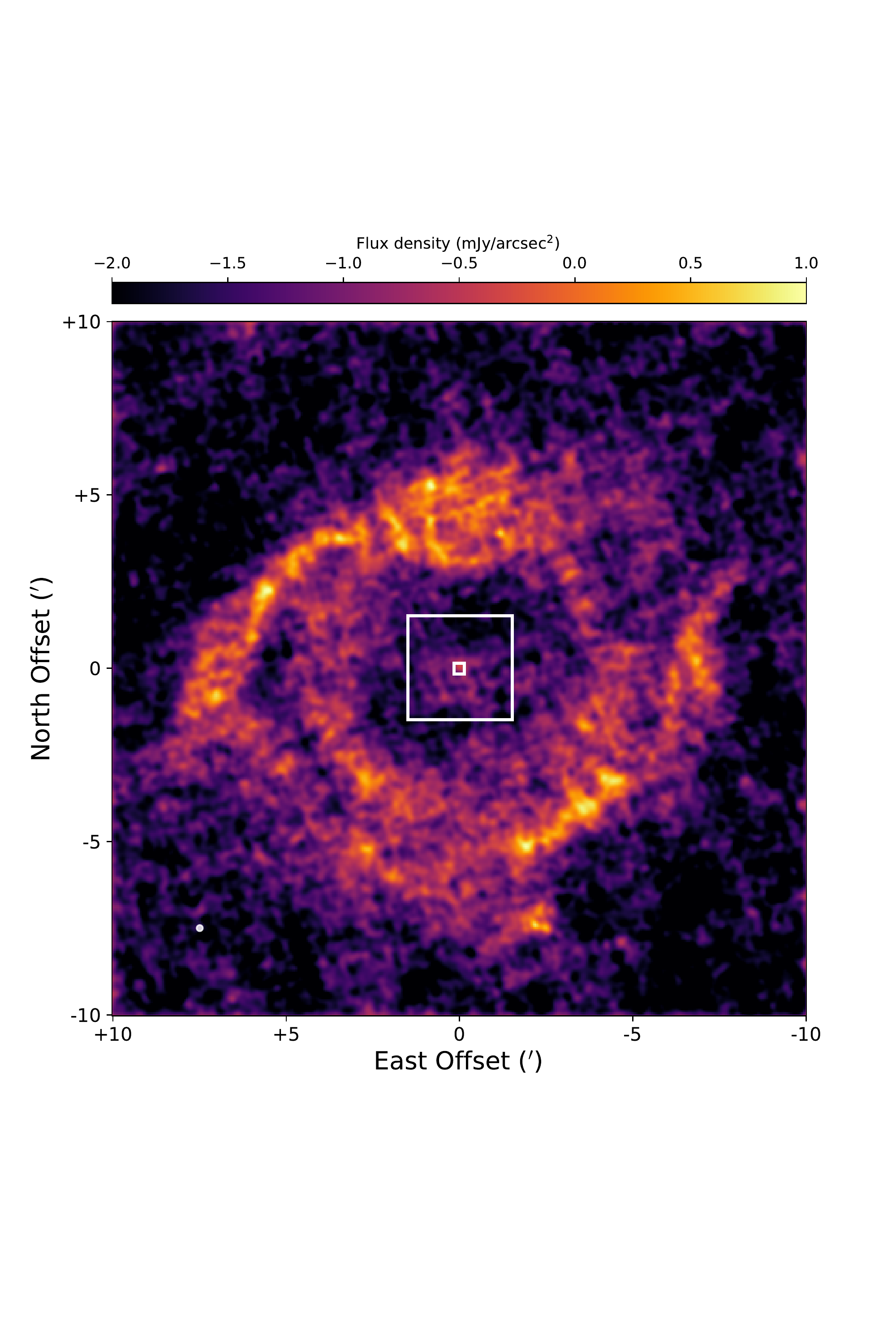}
    \includegraphics[width=0.45\textwidth,trim={0cm 0cm 0cm 1cm},clip]{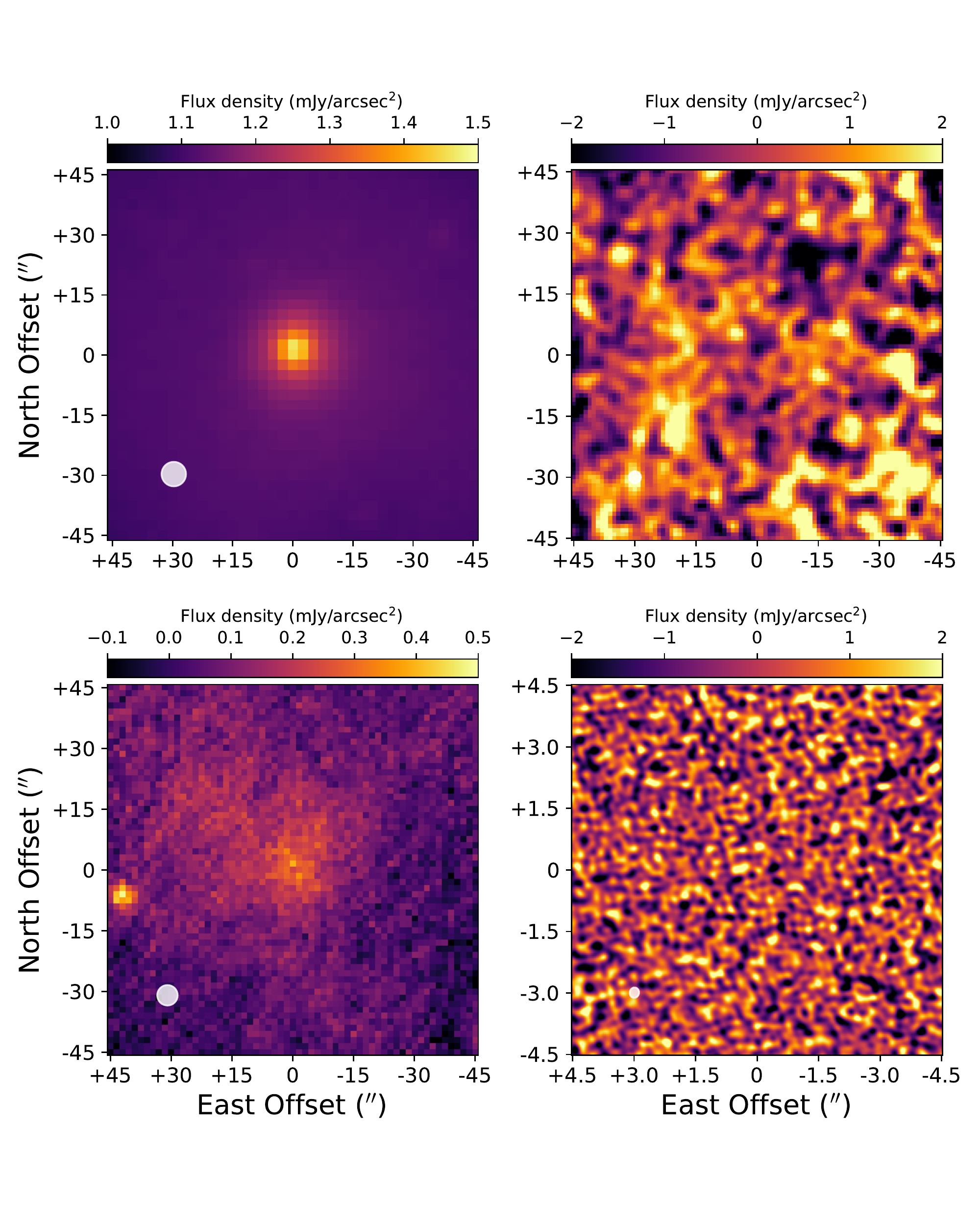}
    \caption{Imaging observations centred on WD~2226-210. \textit{Left}: The Helix Nebula observed at 70~$\mu$m by \textit{Herschel}/PACS \citep{2015VandeSteene}. The image has been smoothed with a Gaussian kernel with FWHM 5\farcs4 to enhance its appearance. The region centered on WD~2226-210 and covered by the four cutaway panels to the right is denoted by the white boxes. \textit{Right}: The top row is \textit{Spitzer}/MIPS 24~$\mu$m {\citep{2007Su}} (left) and SOFIA/HAWC+ 54~$\mu$m (right), and the bottom row is \textit{Herschel}/PACS 70~$\mu$m (left) and ALMA Band 6 (right). Only the \textit{Spitzer} and \textit{Herschel} observations detect the excess emission associated with the white dwarf. The orientation of all images is north up, east left. The instrument beam for each observation is denoted by the white shaded ellipse in the bottom left corner of the panel.}
    \label{fig:helix_all}
\end{figure*}

Here we provide details of the new observations of the Helix Nebula we have obtained with SOFIA and ALMA, used to determine the brightness and spatial extent of the white dwarf's infrared excess emission. We present the imaging observations most relevant to this work, spanning mid-infrared to millimetre wavelengths, in Figure \ref{fig:helix_all}.  A summary of the flux density measurements used to scale the white dwarf photosphere model and to fit the disc spectral energy distribution (SED), combining these new data with existing archival measurements, are provided in Table \ref{table:helix_phot}.  {In this table we break down the excess flux measurements of the target into three categories, `Total', `Extended', and `Compact'. `Total' flux measurements are a combination of the observed flux (or upper limits) associated with both a point-like source and extended emission associated with the white dwarf position. `Compact' flux measurements denote detection of a point source (or 3-$\sigma$ upper limit) associated with the white dwarf position. `Extended' measurements for \textit{Herschel} are related to the presence of a broad feature centred to the NW of the white dwarf with an angular extent of $28\arcsec\times24\arcsec$, whilst the `Extended' measurement for ALMA references the 3-$\sigma$ limit on the presence of a face-on, spatially resolved disc 3 beams in diameter.} The reduction and analysis summarised here is not a complete and detailed overview of the process. The exact reduction steps used can be inferred from the scripts available, along with the reduced data, analysis scripts, and models, in a GitHub repository\footnote{\href{https://github.com/jontymarshall/Formation_of_the_Helix_Nebula}{https://github.com/jontymarshall/Formation\_of\_the\_Helix\_Nebula}}.

SOFIA/HAWC+ Band A (53~$\mu$m) observations centred on WD~2226-210 were taken on 2017 Oct 17 for program 05\_0054 (PI: S. Ertel). The observation was downloaded as a level 3 data product from the SOFIA archive\footnote{SOFIA archive}. The Lissajous map covers a roughly rectangular area 1\farcm5$\times$2\arcmin\, around the white dwarf, with an r.m.s. of 37 mJy/beam for a total integration time of 27 minutes.  {The instrument PSF is 3$\arcsec$ FWHM, equivalent to a spatial resolution of 600~au at the distance of the Helix Nebula.}

Deep \textit{Herschel}/PACS small map observations of the region around the Helix Nebula's white dwarf (PID: OT1\_ksu\_2, PI: K. Su) were taken using the 70 and 160~$\mu$m waveband combination. These unpublished, archival observations were obtained as level 2.5 (pipeline reduced, mosaicked) data products from the \textit{Herschel} Science Archive\footnote{The \textit{Herschel} Science Archive can be accessed at\\ \href{http://archives.esac.esa.int/hsa/whsa/}{http://archives.esac.esa.int/hsa/whsa/}.} \citep{2019Verdugo}. For the analysis presented here we adopt the maps generated using the JScanamorphos algorithm as most reliable for the retention of extended structure. The map has non-uniform coverage, centered on the source, with a rectangular extent of $3\arcmin\times1\farcm5$\, and an r.m.s. of 3~mJy within the central 1\arcmin\, radius of the map where coverage is highest and approximately uniform.  {The instrument PSF at 70~$\mu$m is 5.4$\arcsec$ FWHM, equivalent to a spatial resolution of 1\,000~au at the distance of the Helix Nebula.}

ALMA Band 6 observations of WD~2226-210 were obtained in Cycle 3 (PID: 2015.1.00762.S, PI: S. Ertel) and Cycle 4 (PID: 2016.1.00608.S, PI: S. Ertel). These data sets were obtained from the ESO ALMA Science Archive\footnote{The ALMA science archive can be accessed at\\ \href{http://almascience.eso.org/aq/}{http://almascience.eso.org/aq/}}. In both cases the setup consists of four spectral windows providing a total of 7.5~GHz bandwidth to study the target emission. For the Cycle 3 (C3) observations two windows measure the continuum whilst the third and fourth windows cover the $^{12}$CO (2-1) and $^{29}$SiO (5-4) lines. For the Cycle 4 (C4) observations, three windows lie over the continuum whilst the fourth covers the $^{12}$CO (2-1) line.  Calibration and reduction of the ALMA observations were carried out in CASA using the appropriate CASA version (4.5.3 in C3 and 4.7.6 in C4). Image reconstruction was carried out using scripts supplied by the observatory, adopting Briggs weighting as a compromise between signal-to-noise and resolution. Both data sets had comparable spatial resolution, with a beam FWHM $0\farcs25$ along the major axis (50~au at 200~pc), sufficient to spatially resolve a Solar-system scale disc around the white dwarf. In the C3 data, the measured continuum sensitivity was 9~$\mu$Jy with 3-$\sigma$ line sensitivities of 1.38~mJy/km/s for $^{12}$CO (2-1) and 1.53 mJy/km/s for $^{29}$SiO (5-4). In the C4 data, the continuum sensitivity was 11~$\mu$Jy with a 3-$\sigma$ line sensitivity of 1.74~mJy/km/s for $^{12}$CO (2-1).

 {In summary, we have obtained new images of the central regions of the Helix Nebula centred on the white dwarf with SOFIA/HAWC+ at 53~$\mu$m and ALMA in Band 6 ($\simeq1\,300~\mu$m). We complement these observations with previous archival imaging from \textit{Spitzer}, presented in \cite{2007Su}, and \textit{Herschel}, presented in \cite{2015VandeSteene}. The \textit{Spitzer} data show a compact source at infrared wavelengths (from $8~\mu$m to 70~$\mu$m) centred on the white dwarf position. The \textit{Herschel} far-infrared observations (70 and 160~$\mu$m) from Van de Steene et al. likewise show compact emission associated with the white dwarf, but no emission is detected at sub-millimetre wavelengths (250 to 500~$\mu$m). The archival \textit{Herschel} far-infrared observations (also 70 and 160~$\mu$m), which are deeper than those of Van de Steene et al., show compact emission associated with the white dwarf and its environment consistent with the \textit{Spitzer}/MIPS 70~$\mu$m observation. The new SOFIA/HAWC+ 53~$\mu$m observation has a higher angular resolution than the previous \textit{Spitzer} and \textit{Herschel} 70~$\mu$m observations, but has much lower sensitivity and is a non-detection for the expected excess emission. The ALMA millimetre-wavelength observations are likewise non-detections in both continuum and line emission. In the next section we will combine the available detections and upper limits to model the excess emission and infer its likely origin}. 

\begin{deluxetable}{cccc}
%\centering
\tablewidth{0.45\textwidth}
\tablecolumns{4}
\tablecaption{Summary of photometric measurements. Upper limits are 3-$\sigma$. \label{table:helix_phot}}
\tablehead{
\colhead{Wavelength} & \colhead{Flux density}  & \colhead{Instrument/} & \colhead{Ref.} \\
\colhead{($\mu$m)}   & \colhead{(mJy)} & \colhead{Filter}      & \colhead{}     
}
\startdata
0.44 & 21.6~$\pm$~1.3 & Johnson $B$ & 1\\
0.55 & 14.7~$\pm$~0.4  & Johnson $V$ & 1\\
1.235 & 2.95~$\pm$~0.07 & 2MASS $J$ & 2\\
1.662 & 1.64~$\pm$~0.06 & 2MASS $H$ & 2\\
2.159 & 1.01~$\pm$~0.08 & 2MASS $K_{\rm s}$ & 2\\
3.4 & 0.42~$\pm$~0.02 & \textit{WISE} W1 & 3\\
3.6 & 0.374~$\pm$~0.019 & \textit{Spitzer}/IRAC & 4\\
4.5 & 0.241~$\pm$~0.024 & \textit{Spitzer}/IRAC & 4\\
4.6 & 0.30~$\pm$~0.02 & \textit{WISE} W2 & 3\\
5.8 & 0.171~$\pm$~0.026 & \textit{Spitzer}/IRAC & 4\\
11.6 & 0.9~$\pm$~0.1 & \textit{WISE} W3 & 3\\
22.1 & 73~$\pm$~9 & \textit{WISE} W4 & 3\\
\hline
\multicolumn{4}{c}{Total}\\
\hline
70.0 & 224~$\pm$~33 & \textit{Spitzer}/MIPS & 4\\
70.0 & 258~$\pm$~13 & \textit{Herschel}/PACS & 5\\
70.0 & 239~$\pm$~26 & \textit{Herschel}/PACS & 6\\
160.0 & $<$ 711 & \textit{Spitzer}/MIPS & 4\\
160.0 & $<$ 405 & \textit{Herschel}/PACS & 5\\
250.0 & $<$ 180 & \textit{Herschel}/SPIRE & 6\\
1300.0 & $<$ 0.030 & ALMA Band 6 & 5\\
\hline
\multicolumn{4}{c}{Compact}\\
\hline
8.0 & 0.174~$\pm$~0.017 & \textit{Spitzer}/IRAC & 4\\
24.0 & 48.4~$\pm$~7.3 & \textit{Spitzer}/MIPS & 4\\
54.0 & $<$ 111 & SOFIA/HAWC+ & 5\\
70.0 & 36.0~$\pm$~4.5 & \textit{Herschel}/PACS & 5\\
160.0 & $<$ 45 & \textit{Herschel}/PACS & 5\\
1300.0 & $<$ 0.030 & ALMA Band 6 & 5\\
\hline
\multicolumn{4}{c}{Extended}\\
\hline
70.0 & 222~$\pm$~12 & \textit{Herschel}/PACS & 5\\
160.0 & $<$ 360 & \textit{Herschel}/PACS & 5\\
1300.0 & $<$ 0.162 & ALMA Band 6 & 5\\
\enddata
\raggedright

\tablerefs{1. \cite{2007Harris}; 2. \cite{2006Skrutskie}; 3. \cite{2010Wright}; 4. \cite{2007Su}; 5. This work; 6. \cite{2015VandeSteene}.}

\end{deluxetable}

\section{Modelling}
\label{sec:meth}

Here we summarise the modelling approach used to constrain the spatial extent and density distribution of the circumstellar dust around the white dwarf and determine the dust properties based on the spatial constraints from multi-wavelength imaging and source SED. Once we have an understanding of the dust emission, we then apply a radiative transfer model to determine the minimum dust grain size and size distribution

\subsection{Image analysis and modified blackbody fitting}

Initially, we examine the assembled imaging observations of the Helix Nebula to determine the degree of excess emission as a function of wavelength, and its spatial distribution{; these observations are presented in Figure \ref{fig:helix_all}}. We measure the extent of the extended disc using imaging observations at 24, 54, 70, and 1300~$\mu$m.  From the 24~$\mu$m \textit{Spitzer} observations of \citet{2007Su}, the spatial extent of the proposed circumstellar disc is 30 to 100~au, spanning an angular \emph{diameter} of ~ 0.25$^{\prime\prime}$ to 1$^{\prime\prime}$; this component is therefore unlikely to be spatially resolved in the SOFIA and \textit{Herschel} far-infrared observations. We therefore consider that the system may comprise two components: a compact (point-like) source, and an extended component. The compact component is modelled as a point source using a wavelength appropriate PSF model, whilst the extended component is modelled as a 2D Gaussian profile. The convolved model is then subtracted from the observation and significant residuals identified in the image.

We are aware that in the ALMA observations the point-like component may itself be spatially resolved, and therefore the disc orientation (surface brightness) becomes important as a constraint at millimetre wavelengths. A broad extended component may remain undetected by ALMA due to either low surface brightness, or insensitivity to the appropriate angular scale due to the interferometer configuration. Furthermore, we also consider the presence of (spatially resolved) gas emission at millimetre wavelengths from the system. Two common species, CO and SiO, were covered by spectral windows in the ALMA observations. These lines are diagnostic of emission from post-AGB envelopes and icy planetesimal belts. Both CO and SiO emission might be expected from a remnant post-AGB disc, whereas volatile rich planetesimals may leave a detectable CO emission in a debris disc. The geometry of the line emission would further inform our understanding of the origin of the excess emission. 

We then proceed to model the SED. We first model the emission as a combination of modified blackbodies specific to the number of spatial components in the system obtained in the previous step \citep[e.g.][]{2008Wyatt}. The modified blackbodies are defined by temperature $T_{\rm bb}$, break wavelength $\lambda_{0}$, and exponent $\beta$, such that

\begin{equation}
    F(\lambda) \propto 
\begin{cases}
    B(\lambda,T_{bb}), & \lambda \leq \lambda_{0}\\
    B(\lambda,T_{bb})\times(\lambda_{0}/\lambda)^{\beta}, & \lambda > \lambda_{0}
\end{cases}
\end{equation}

with the following additional constraints:
\begin{itemize}
    \item [(a)] The rising \textit{Spitzer}/IRS spectrum at mid-infrared wavelengths is entirely accounted for by the compact component.
    \item [(b)] The temperature of the compact component, being strongly constrained by the \textit{Spitzer}/IRS spectrum \citep{2007Su}, is fixed as 102~K and the temperature of the extended component must be lower than the compact component.
    \item [(c)] The break wavelength ($\lambda_{0}$) for both components must be at least the shortest wavelength for which a detection exists (i.e. 24~$\mu$m for the compact component and 70~$\mu$m for the extended component).
    \item [(d)] The beta exponents for both components must lie between 0 and 4, which covers the range observed for debris discs. The beta exponent may be different for the two components.
\end{itemize}
We determine the best fit parameters for all components simultaneously. To do this we create 10,000 sets of values drawn from the appropriate ranges for each parameter ($T_{bb}$, $\lambda_{0}$, $\beta$) and scale the resultant modified blackbodies to the measured flux density of the system. Models that violated the upper limits for either component were discarded from the final ensemble. For the remainder that were consistent with the available data, we calculated the best-fit values for the two components in combination through error weighted least-squares fitting, and the mean and standard deviation of the valid parameter sets in isolation.

\subsection{Radiative transfer modelling}

We then calculate the appropriate dust grain size range and distribution necessary to match the shape of the excess emission fitted by the modified blackbodies. The analytical radiative transfer model uses physical constraints from the central white dwarf luminosity to calculate emission from a dusty circumstellar envelope (or disc, or both) and infer the dust grain size range and size distribution. We assume the dust emission is optically thin, consistent with the observed fractional luminosity at infrared wavelengths ($L_{\rm dust}/L_{\rm WD} \simeq 10^{-3}$), and that the dust size distribution is defined by a power law distribution with minimum and maximum grain sizes, and the slope of the size distribution between those limits. The spatial distribution of the dust emission, constrained in extent by the imaging data, is assumed to be a disc with an inner radius $R_{\rm in}$ and outer radius $R_{\rm out}$, with a power law slope in surface brightness $\alpha$, such that the surface brightness $\propto (R/R_{in})^{\alpha}$. Additional component(s) of dust emission will be included in the modelling with the same underlying structural model (but different limits) to satisfy the observed spatial distribution of emission from the system.

\section{Results}
\label{sec:resl}

Here we present our findings regarding the structure and emission properties of the Helix white dwarf's excess emission, based on the multi-wavelength imaging and photometry data sets. We first consider the extent of the emission region, combining archival mid- and far-infrared imaging from \textit{Spitzer} and \textit{Herschel} with new far-infrared SOFIA/HAWC+ and millimetre ALMA data. We then determine the spectral shape of the excess emission, informed by infrared to millimetre wavelength photometry and upper limits. We then combine these constraints using a radiative transfer model to infer the dust grain properties necessary for consistency between them.

\subsection{Imaging and spectral energy distribution}

\begin{figure*}
    \centering
    \includegraphics[width=\textwidth]{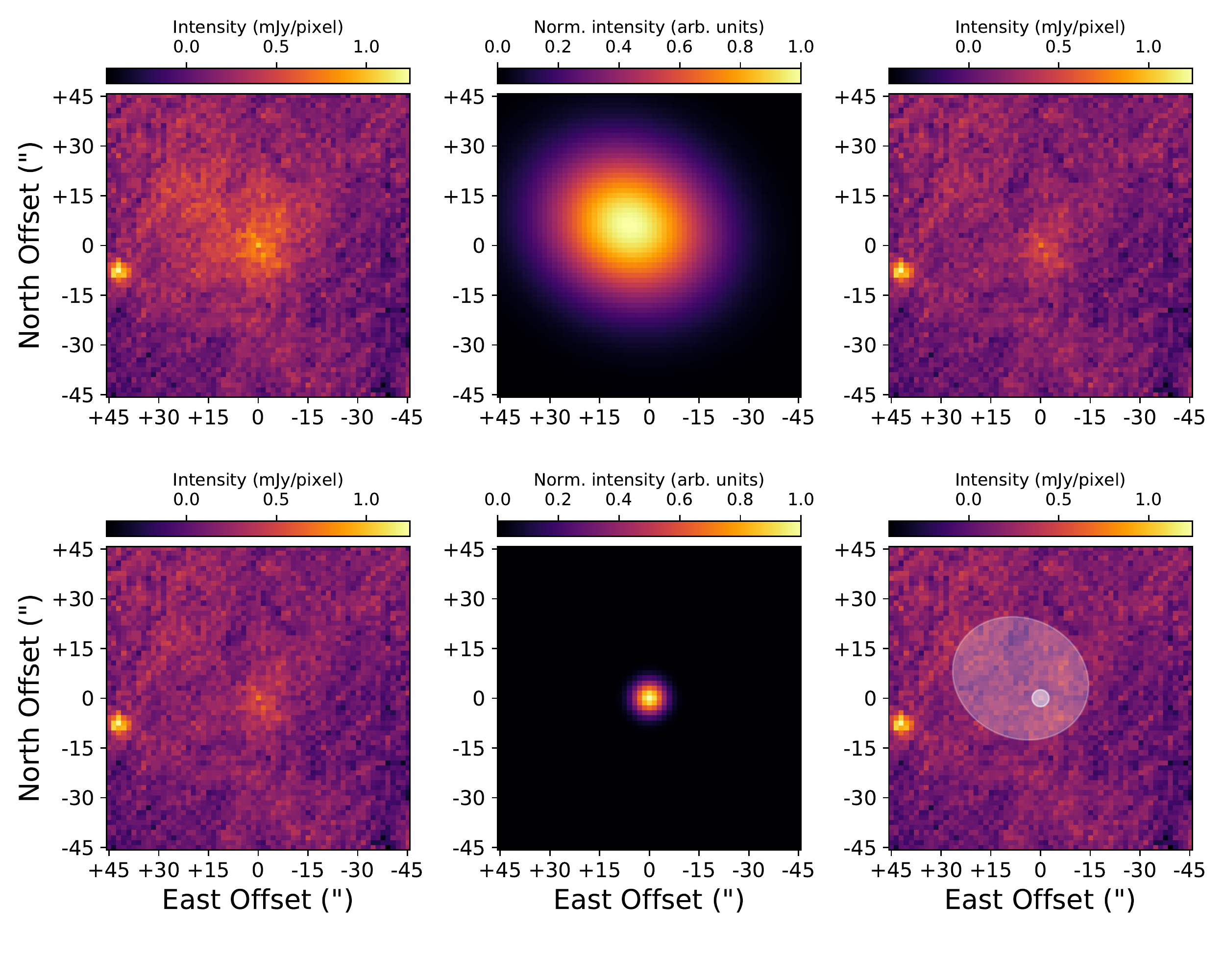}
    \caption{\textit{Herschel}/PACS 70~$\mu$m imaging observations of the central white dwarf of the Helix Nebula. The series of images, from top left to bottom right, show the results of fitting and subtracting a two component model for the emission centred on the white dwarf position. \textit{Top}: The original observation (left), is followed by the 2D Gaussian model for the extended component (middle) which appears offset from the white dwarf location, and the residuals after subtraction of that component (right). \textit{Bottom}: The residuals from subtraction of the extended component (left) are then fitted with a 2D Gaussian model matching the instrument PSF (FWHM 5\farcs4) (middle), and the final residual map after subtraction of both components are presented (right).  {The shaded ellipses in this panel denote the extent of the extended (FWHM $28\arcsec\times24\arcsec$, $\phi$ $65\degr$) and compact (FWHM $5.4\arcsec$) components subtracted from the image}. The images are oriented north up, east left. The plate scale is 1\arcsec\, per pixel.}
    \label{fig:herschel_pacs}
\end{figure*}

Mid- and far-infrared excess emission close to the Helix central white dwarf was first discovered with \textit{Spitzer}, revealing spatially unresolved emission located at the stellar position that was seen at wavelengths from 8~$\mu$m to 70~$\mu$m \citep{2007Su}, {see Figure \ref{fig:helix_all}}. The emission could be clearly disentangled from the bulk of the nebula’s emission due to its compactness (diameter $< 2\arcsec$ at 200~pc, based on the 8~$\mu$m image). Fitting the  {excess SED} revealed it to have a temperature of $\geq$~86~K and a corresponding size of several 10s au, comparable to the extent of the Edgeworth-Kuiper belt, but it was not possible to constrain the dust properties from the limited wavelength range of the available data.

The SOFIA/HAWC+ Band A (54~$\mu$m) observation does not have any evidence of a point-like source at the white dwarf position, as shown in Figure \ref{fig:helix_all}. At first glance, this is seemingly inconsistent with the previously inferred dust temperature and morphology of the \textit{Spitzer} {imaging observations and photometry}, including the rising \textit{Spitzer}/IRS spectrum up to 35~$\mu$m. Since the \textit{Spitzer}/MIPS and SOFIA/HAWC+ maps have comparable resolution (FWHM $\simeq 6\arcsec$), the non-detection in the longer wavelength observation places a strong constraint on the total brightness of the excess emission, despite its limited sensitivity (r.m.s. 37~mJy). We can therefore infer that within 600~au of the white dwarf, the total emission must be under 120~mJy.

In contrast, archival \textit{Herschel}/PACS 70~$\mu$m imaging data (PI: K. Su) surprisingly show extended emission associated with the white dwarf's position. That emission can be decomposed into two components, which we have fitted using a PSF model for the compact component, with a flux density of  36~$\pm$~4.5~mJy,  and a 2D Gaussian for the extended emission, with a flux density of 222~$\pm$~12~mJy, as shown in Figure \ref{fig:herschel_pacs}. The brightness of both components are consistent with the non-detection in the SOFIA/HAWC+ 54~$\mu$m image due to the much greater sensitivity of \textit{Herschel}/PACS at 70~$\mu$m. Furthermore, the level of excess emission from the extended component and its non-detection in sub-millimetre \textit{Herschel}/SPIRE maps of the wider nebula \citep{2015VandeSteene} rules out the presence of a substantial mass of cold dust at larger distances from the white dwarf which would have been undetected in the SOFIA/HAWC+ map due to dilution across multiple beams.

The ALMA observations of the Helix white dwarf are also non-detections, both in continuum and line measurements. The largest angular scale for the array configuration was about $4\arcsec$ which rendered the ALMA observations blind to the extended component seen in the \textit{Herschel}/PACS map. However, these maps are deep, with a continuum r.m.s. around 10~$\mu$Jy. The compact component of the emission should have been easily recoverable in the observations given its presumed extent based on the shorter wavelength maps, if it followed a spectral slope consistent with debris disc emission (i.e. including large grains up to mm sizes). Therefore, the ALMA observations provide a strong constraint on the surface brightness of the excess emission at millimetre wavelengths.  {The r.m.s. sensitivity of the line observations were 0.38 mJy/km/s for CO (2-1) and 0.51 mJy/km/s for SiO (5-4). Assuming a line width of 10 km/s (typical for debris discs), optically thin emission, and local thermodynamic equilibrium for the gas ($T_{\rm gas} = 100~K$), we derive 5-$\sigma$ upper limits to the integrated line fluxes of 2.3$\times10^{-17}$ W/m$^{2}$ for SiO (5-4) and 3.1$\times10^{-17}$ W/m$^{2}$ for SiO (5-4). These limits are equivalent to gas masses of 1.2$\times10^{-4}~M_{\oplus}$ in CO and 2.7$\times10^-2~M_{\oplus}$ in SiO.} The non-detection of CO or SiO line emission from the white dwarf rules out the presence of substantial gas mass, contrary to expectations if the excess emission originated from a remnant post-AGB disc.

\begin{figure}
    \centering
    \includegraphics[width=0.45\textwidth]{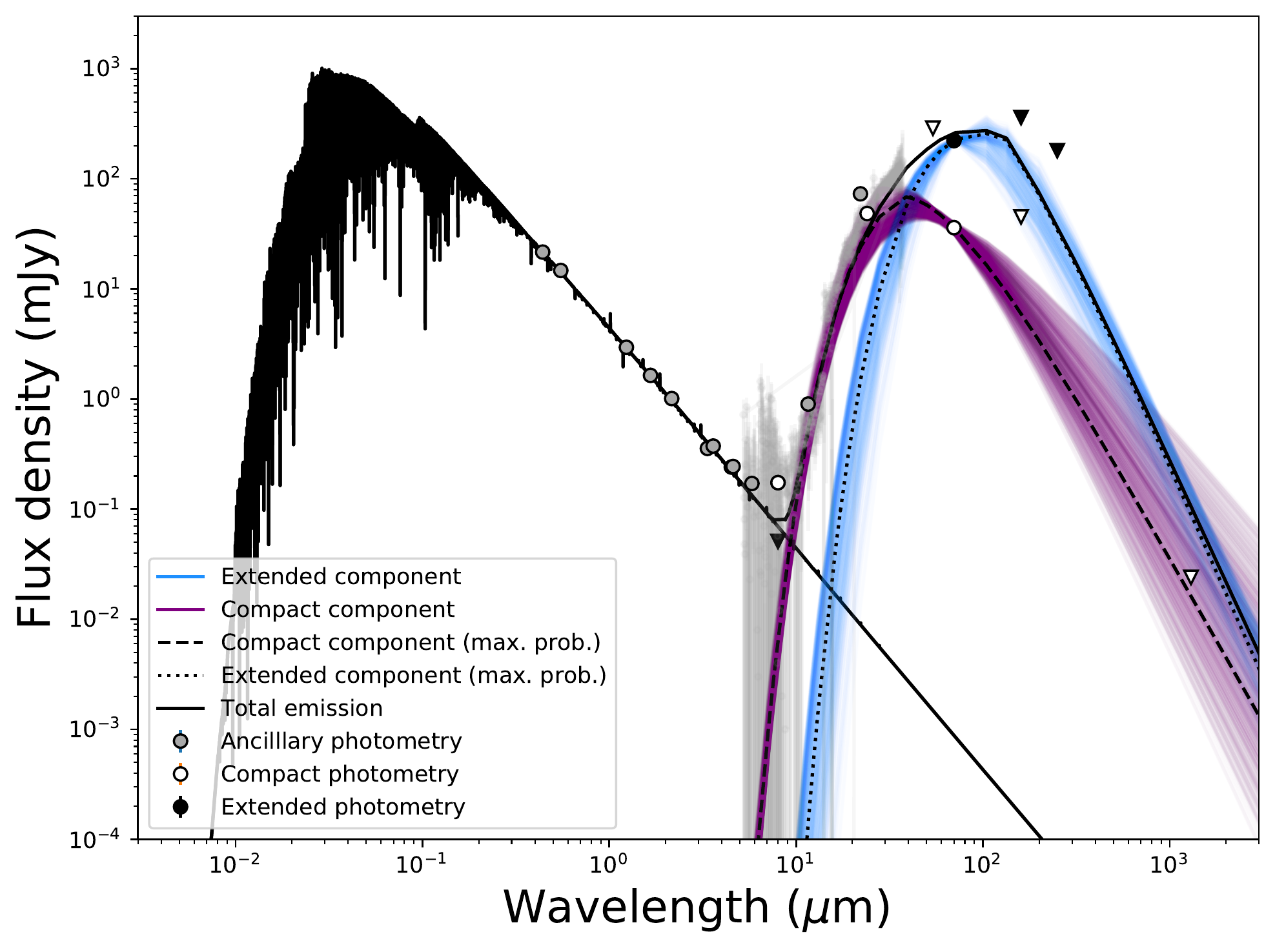}
    \caption{SED of the central source of the Helix Nebula, WD~2226-210. The black solid line denotes the white dwarf photopshere model scaled to optical and near-infrared photometry, shown as grey circles. The black dashed lines denote the best-fitting modified blackbody models representing the contribution of the compact and extended dust components to the total emission. Purple and blue lines denote models consistent with the observations drawn at random from the tested parameter ranges for the compact and extended components, respectively. White circles and triangles denote measurements and upper limits to the compact component, whilst the black circles and triangles denote the same for the extended component. Upper limits are 3-$\sigma$, taking into account the source extent. The grey line denotes the \textit{Spitzer}/IRS spectrum and its associated uncertainty.  \label{fig:helix_sed}}
\end{figure}

From the SED presented in Figure \ref{fig:helix_sed} we see that the shape of the compact component (purple) is strongly constrained by both the \textit{Spitzer}/IRS spectrum and the \textit{Herschel}/PACS photometry. In conjunction, these observations confine the shape of the excess to being a sharp rise followed by a sharp fall-off. The SOFIA HAWC+ upper limit at 54~$\mu$m further constrains the total brightness of the excess such that the peak of emission must occur before that wavelength. However, the non-detection at (sub-)millimetre wavelengths of the compact component leaves its spectral slope unconstrained, with the ALMA measurements offering only weak constraints on the shape if we assume the dust emission region would be unresolved (consistent with a compact Kuiper belt analogue around the white dwarf). By contrast, the extended component (blue) is only weakly constrained with a single measurement in the \textit{Herschel}/PACS 70~$\mu$m map. Attributing the detected excess at shorter wavelengths to the compact component alone provides some restriction on the temperature of the extended component, whilst the weak upper limits in the sub-millimetre and millimetre from \textit{Herschel} SPIRE and ALMA enable us to rule out a massive cold component to the total emission. The best fit parameters of the modified blackbody models used to interpret the SED are given in Table \ref{table:helix_sed}. 

\begin{deluxetable}{lcc}
\tablewidth{0.45\textwidth}
\tablecolumns{3}
\tablecaption{Summary of SED model fitting. The temperature of the compact component was held fixed at 102~K. \label{table:helix_sed}}
\tablehead{
\colhead{Parameter} & \colhead{Compact}  & \colhead{Extended}
}
\startdata
Temperature (K)        & 102       & 50~$\pm$~6 \\
$\lambda_{0}$ ($\mu$m) & 25~$\pm$~7 & 96~$\pm$~29 \\
$\beta$                & 1.4~$\pm$~0.7 & 1.5~$\pm$~0.2 \\
\enddata
\end{deluxetable}

The available data offer no real constraint on the nature of the extended component seen in the \textit{Herschel} 70~$\mu$m map with weak upper limits either side of that to constrain its behaviour. Multiple interpretations could fit the evidence equally well, such as it being a blow-out halo from the compact component, part of the wider Helix nebula coincidentally superimposed upon the white dwarf, or a diffuse remnant of the dispersing post-AGB envelope. However, the compact component extent is relatively well constrained to within a few 100~au of the white dwarf and its SED is surprisingly well constrained with a steep rise, quick turnover, and steep fall off. In the next sub-section we perform a further analysis of the compact component, applying these constraints in a radiative transfer model.

\subsection{Radiative transfer modelling}

\begin{figure*}
    \centering
    \includegraphics[width=0.45\textwidth]{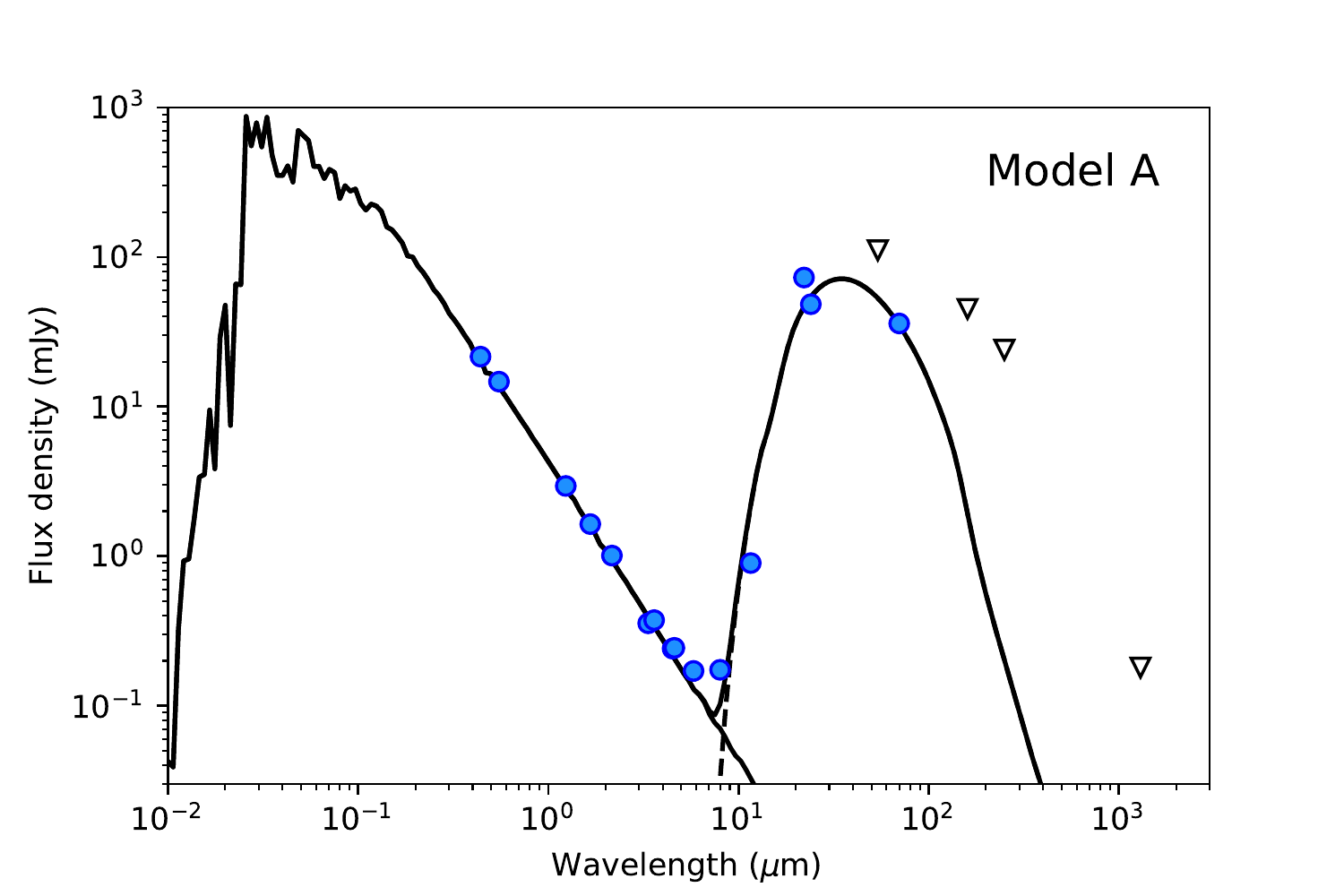}
    \includegraphics[width=0.45\textwidth]{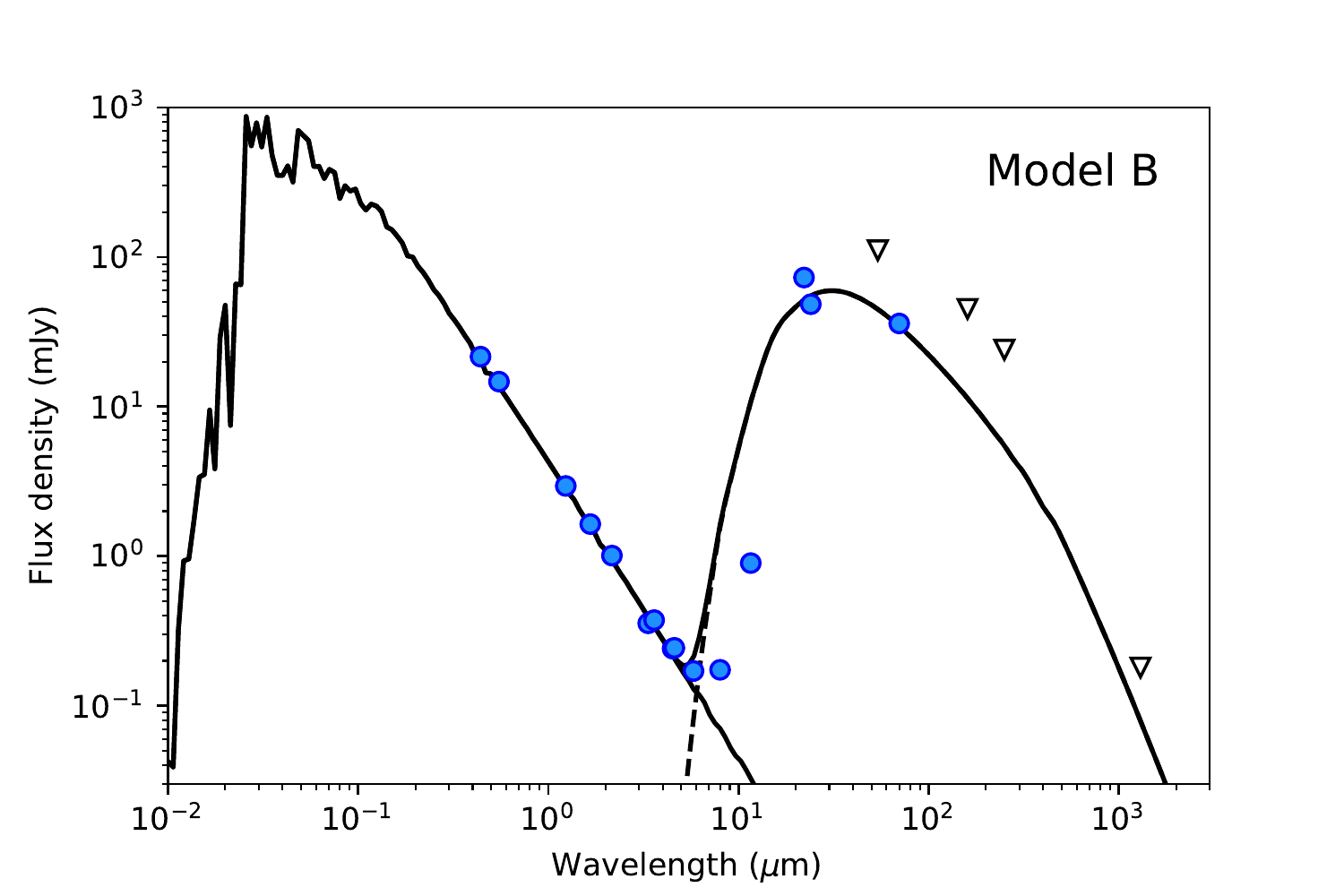}\\
    \includegraphics[width=0.45\textwidth]{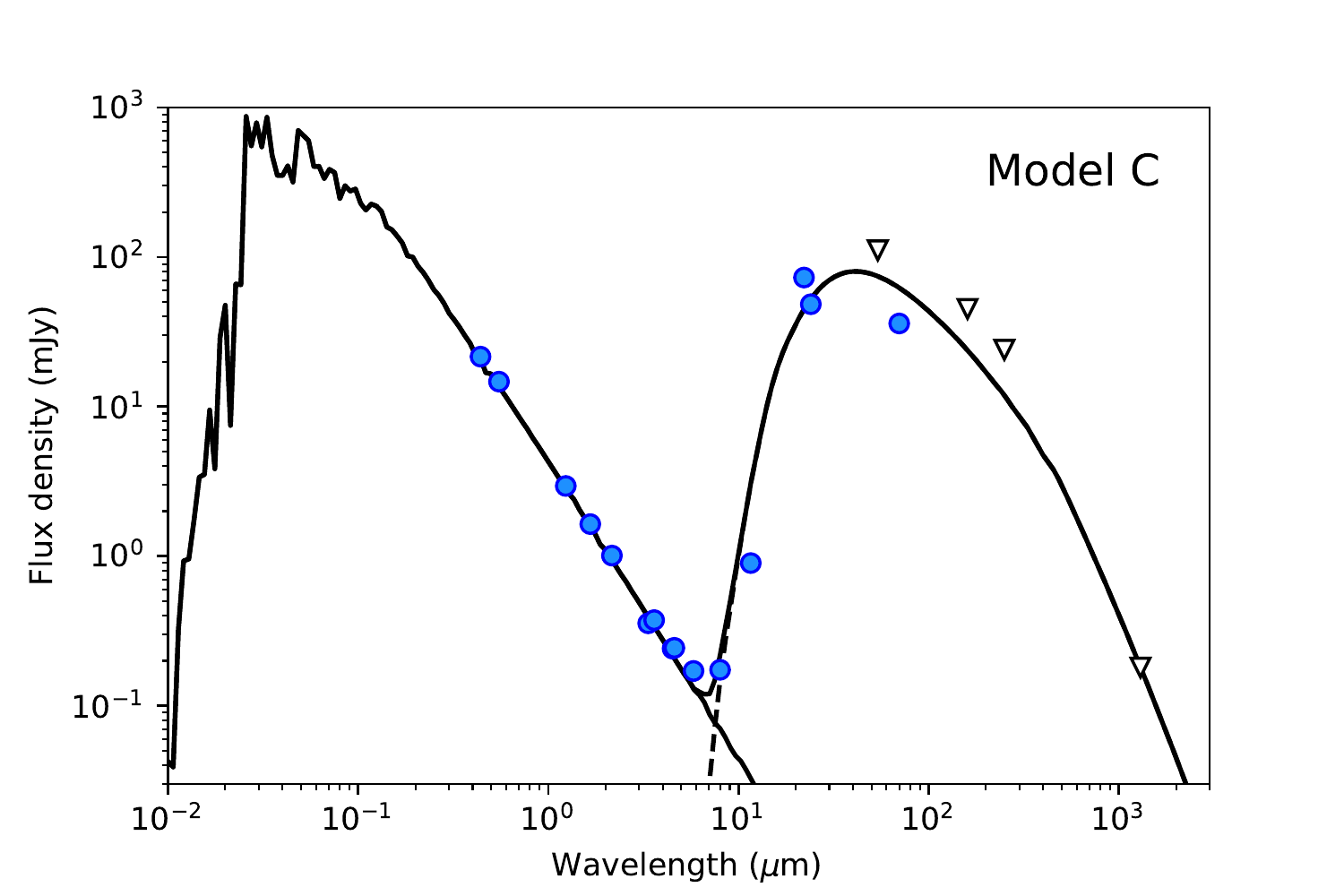}
    \includegraphics[width=0.45\textwidth]{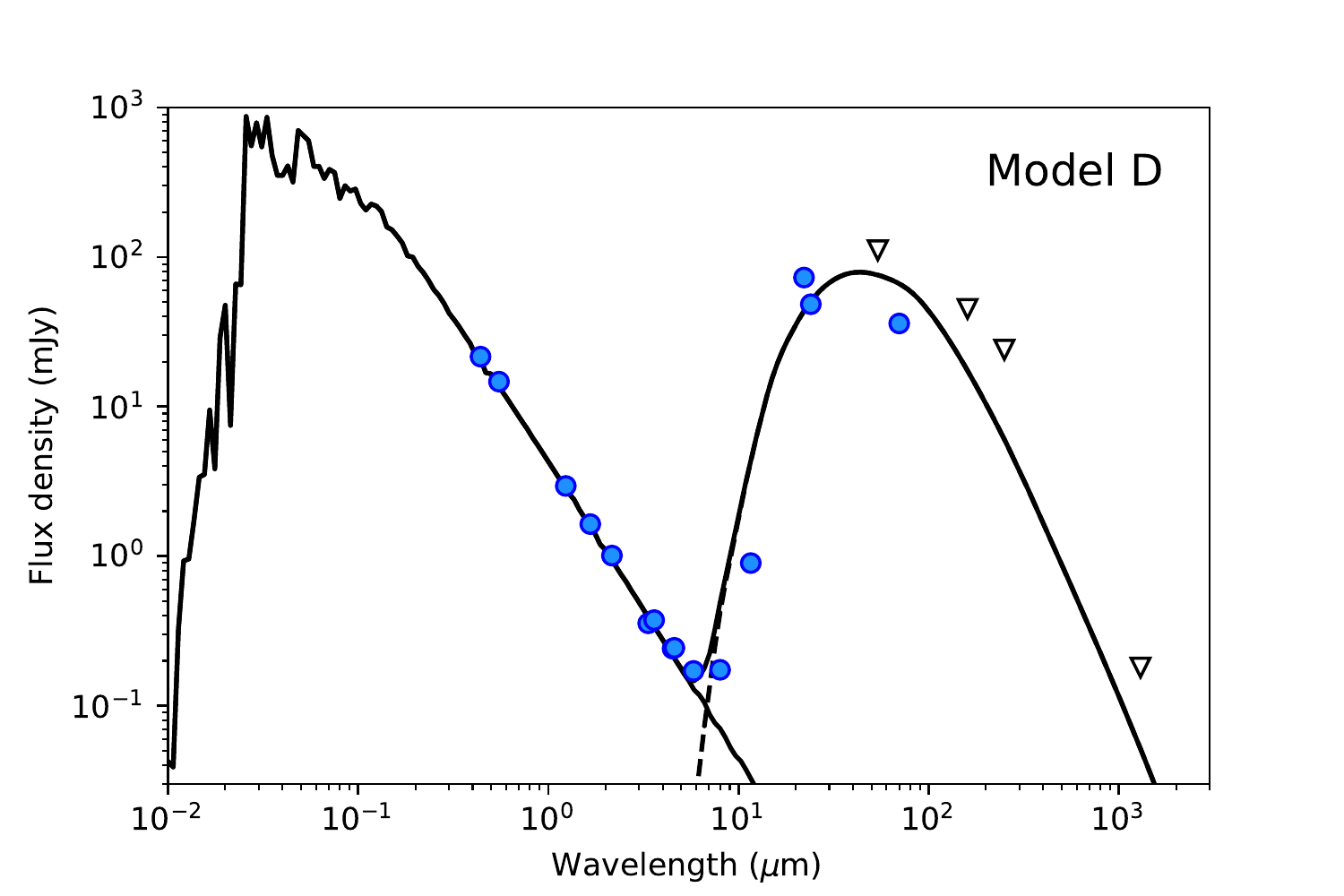}\\
    \caption{A comparison of observations to model SEDs illustrating four scenarios consistent with the presence of circum-white dwarf excess emission in the Helix Nebula. Photometry from Table \ref{table:helix_phot} are presented as blue data points, upper limits (3-$\sigma$) are downward pointed triangles. \textit{Top left}: In scenario A, we consider a wide annulus at 60 to 80~au with dust grains ranging in size from 2 to 20~$\mu$m. \textit{Top right}: In scenario B, we consider a compact Solar system-scale debris disc with grains ranging in size from 60~$\mu$m (blowout limit) to 1~mm. \textit{Bottom left}: In scenario C, we consider an extended debris disc with grains ranging in size from 60~$\mu$m (blowout limit) to 1~mm. \textit{Bottom right}: In Scenario D, we consider an extended debris disc with grains ranging in size from 10~$\mu$m (1/6th blowout limit) to 1~mm. Of the four scenarios considered here, only scenario A, with a small minimum grain size and narrow size distribution is consistent with all the observations. Limiting the minimum size of dust grains to the blowout limit and above clearly violates the steep mid-infrared rise and/or the far-infrared and millimetre constraints, for either a compact or an extended disc architecture.}
    \label{fig:comparison_seds}
\end{figure*}

\begin{deluxetable*}{lcccc}
%\centering
\tablewidth{\textwidth}
\tablecolumns{5}
\tablecaption{Summary of model parameters used to calculate SEDs presented in Figure \ref{fig:comparison_seds}. \label{tab:comparison_models}}
\tablehead{
\colhead{Parameter}   & \colhead{Model A} & \colhead{ Model B} & \colhead{Model C} & \colhead{Model D}    
}
\startdata
Disc inner radius, $R_{\rm in}$ (au) & 60 & 30 & 30 & 30 \\
Disc outer radius, $R_{\rm out}$ (au) & 80 & 50 & 90 & 90 \\
Exponent of surface brightness profile, $\alpha$ & 0.0 & -1.0 & -1.0 & -1.0 \\
Minimum grain size, $a_{\rm min}$ ($\mu$m) & 2 & 60 & 60 & 10 \\
Maximum grain size, $a_{\rm max}$ ($\mu$m) & 20 & 1000 & 1000 & 1000 \\
Exponent of size distribution, $\gamma$ & 3.5 & 3.5 & 3.5 & 3.5 \\
$M_{\rm dust}$ ($\times10^{-3}~M_{\oplus}$) & 8 & 10 & 10 & 10 \\
\enddata
\raggedright
%\tablerefs{}
\end{deluxetable*}

We used standard debris disc modelling tools to fit the far-infrared SED of the compact excess \citep{2012Ertel} and simulate spatially resolved observations \citep{2011Ertel}. This is appropriate to determine the spatial and grain size distribution of any accumulation of optically thin circumstellar dust (the fractional luminosity of the Helix excess suggests it is optically thin). For each model fit to the SED, a spatially resolved ALMA image of the emission was simulated and compared to the ALMA upper limit. We used simple power-law radial dust and grain size distributions and realistic dust emissivities commonly employed in debris disc modelling. A broad parameter space of the spatial and grain size distribution was explored, as illustrated in Figure \ref{fig:comparison_seds}, and the exact model parameters for each scenario are summarised in Table \ref{tab:comparison_models}.

The lower limit of the grains size derived from SED modelling of debris discs around luminous main sequence stars is similar to the blow-out size (i.e. the minimum size for which dust grains can remain in a bound orbit around the host star under the influence of radiation pressure) of the dust grains inferred from the stellar properties \citep[see sect. 4.9.1,][]{2020Horner}. For the Helix white dwarf ($T_{\rm WD} = 103\,600~\pm~5\,500$~K, $L_{\rm WD} = 67.6~L_{\odot}$), this size is 60~$\mu$m. Such dust grain sizes are not consistent with the steep slope of the far-infrared excess emission toward longer wavelengths as constrained by the \textit{Herschel} data. The \textit{Herschel} data also rule out that the compact component is too extended to be detected by ALMA. Instead, the ALMA non-detection provides strong constraints on the excess brightness at millimetre wavelengths.

The steep slope of the SED can only be explained by the presence of small, micron sized grains that dominate the emission but have a very low emissivity at long wavelengths \citep{2012Ertel,2016Marshall}. Such a grain size is similar to that of typical dust grains released by evaporating comets in the Solar system \citep{2001Mason}. These grains cannot be on stable orbits around the Helix central star due to their removal by radiation pressure and Poynting-Robertson drag, and must be replenished continuously. The removal of the dust by radiation pressure could also explain the broad extended emission seen in the \textit{Herschel} data, the radial distribution of which is consistent with such a scenario.

\section{Discussion}
\label{sec:disc}

Here we consider the probability, based on the lines of evidence presented above, that the observed excess emission originates from a leftover planetesimal belt, a dissipating cometary cloud, or the remnants of the post-AGB circumbinary disc.

The white dwarf central star WD 2226-210 has a well determined mass of $0.60~\pm~0.02~M_{\odot}$ \citep{2009Benedict}.  Late-type binary companions as cool as late M stars are not detected \citep{1999Ciardulo} and the photometric variability with 2.77 days period detected using the TESS light curves excludes a secondary companion with a mass in the range $0.16~M_{\odot} \le M_{\star} \le 2.5~M_{\odot}$ \citep{2020Aller}. At this period the now substellar secondary must have gone through common envelope evolution.  During the system's post main sequence evolution, any circumbinary planets or planetesimal belts would have been disrupted by the effect of mass loss from the primary, causing orbits to expand outward if located at a distance that avoids tidal engulfment \cite{2007Villaver, 2009Villaver}. Assuming a standard initial-to-final mass relation for the primary the Helix Nebula white dwarf must have a 1.5~\Mso\, stellar progenitor \citep{2009Benedict}. This assumes that the secondary star did not affect the mass-loss rate evolution, which is largely unknown if the system has experienced evolution through a common envelope phase. The engulfment scenario is complex; whether or not a planet is engulfed by the expanding progenitor as it evolves from the main sequence is dependent not only on the progenitor mass and planetary semi-major axis, but also mass of the planetary companion. Different simulations demonstrate engulfment by a 1-2~$M_{\odot}$ progenitor for a gas giant planet with an initial orbital semi-major axis up to 4~au (for a 5~$M_{\rm Jup}$ planet). The orbital distance of the most distant planet that would be engulfed by this star is between 1.7~au (for a terrestrial planet) and 3~au (for a Jupiter-like planet) \citep{2009Villaver,2012Mustill}. Note, however, that planetary systems that host more than one planet are expected to undergo instabilities following post-AGB mass-loss causing planets/planetesimals to be send to orbits close to the WD tidal disruption radius \citep{2018Mustill,2021Maldonado}.

The non-detection of significant gas emission (either CO or SiO) in the ALMA observations of WD~2226-210 places strict constraints on the evolution timescale for any post-AGB binary disc in this system. The Helix Nebula has an estimated total mass of 0.9~\Mso, with an ionized gas mass of 0.3~\Mso\, \citep{1999Henry} and 0.6~\Mso\, in molecular gas \citep{1999Young}. This is a substantial amount of molecular gas, but there is no strong evidence for a remnant disc of material around the central white dwarf. Most of the ${\rm H}_{2}$ is distributed at large angular scales beyond 1\farcm3 up to 5\farcm1 from the central white dwarf \citep{2009Matsuura}. 

Observations of post-AGB disc systems have revealed disc masses in ranging from 8$\times10^{-4}$ up to 10$^{-2}$~\Mso\,\citep{2018Bujarrabal,2021GallardoCava,2022Kluska}. These discs have short lifetimes believed to be around 10$^{4}$ but perhaps up to 10$^{5}$~years \citep{2019Oomen}; the non-detection obtained here is therefore consistent with the shortest expected timescales for the dissipation of the post-AGB circumbinary discs. As a hot, young white dwarf WD~2226-210 exhibits substantial radiation pressure, with a minimum dust grain size of $a_{\rm min} \simeq 60~\mu$m. Dust growth in the post-AGB circumbinary disc could create grains large enough to remain bound to the stellar remnant and withstand this radiation pressure, but dust grains of this size are too large to satisfactorily fit the observed SED. We therefore discard the idea of a post-AGB circumbinary disc as the potential origin for the excess emission.

The size distribution of dust grains in debris discs, produced by the collisional attrition of planetesimals, typically starts around 1 to 10~$\mu$m, at a few times the radiation blowout limit for the host star luminosity \citep{2014Pawellek,2015Pawellek}. The size distribution extends up to millimetre or centimetre sized pebbles, with a (sub-)millimetre size distribution exponent $q$ between 3 and 4 \citep{2016Macgregor,2017Marshall,2021Norfolk}. We find that the spectral slope is constrained by the ALMA upper limits to the top end of this range for compact configurations consistent with a planetesimal belt -- particularly in the edge-on case. For more spatially diffuse emission, the limit on the spectral slope is relaxed, allowing for smaller $q$ values, but then the architecture ceases to be consistent with a planetesimal belt and begins to look more like a shell \citep[e.g.,][]{2018Matra,2021Marshall}. The non-detection of millimetre wavelength emission from the system therefore counts against the planetesimal belt hypothesis for the origin of the observed excess.

  {The incidence of debris discs around main sequence stars of comparable masses to the progenitor star of the Helix Nebula is around 20 to 30~\% \citep{2014Thureau,2017Holland,2018Sibthorpe}. Many young white dwarfs have infrared excesses associated with substantial dust- and gas-rich discs, with an incidence of a few per cent. As the white dwarf cools the brightness of an attendant debris disc likewise drops, especially at infrared wavelengths, confounding detection of any dusty excess for systems beyond a few Myr into the post-main sequence phase of their evolution. Models of the post-main sequence of A star debris discs establish that with current sensitivity constraints one such system within 200~pc would be detectable by current facilities \citep{2010Bonsor}. The Helix Nebula excess identified by \cite{2007Su} is the only example of such a system found to date, and is consistent with the expectations of \cite{2010Bonsor}.}
 
  {The key differences between old white dwarfs and the Helix Nebula's white dwarf are its youth and its much higher luminosity. Around old white dwarfs relatively massive debris discs can persist and are still undetectable because the dust is not heated to any noteworthy degree, this is not the case for any belt around the Helix Nebula's white dwarf \citep{2011Bonsor}.  So it is much harder to hide a belt there, in particular if it is massive enough to produce the huge influx of material into the inner system over the age of the nebula through scattering by a planet. Potentially, we just don’t see a planetesimal belt’s excess because it is on the R-J tail of the warmer dust’s emission and we have few, weak constraints on the total emission from the system at long wavelengths. However, the slope of that emission is already too steep to explain for that dust in a normal debris disc, hence we need dust grains too small for a steady-state debris disc, so it would be hard to hide much cool dust in there.}

 {Steep sub-millimetre slopes have been observed at far-infrared or millimetre wavelengths for a handful of debris disc systems \citep{2012Ertel,2016Marshall}. The nature of the dust in these steep SED debris disc systems is still uncertain. However, in each case the minimum dust grain sizes inferred for those debris disc systems is consistent with the blow out grain size expected from radiation pressure \citep{1979BurnsLamySoter}, which is explicitly not the case here.}

The main constraint that speaks against a debris disc as the origin for the observed excess around the Helix Nebula white dwarf is the small grain size required to fit the shape of the SED from 12 to 70~$\mu$m. The sharp rise, and fall of the SED can only be fit satisfactorily with dust grains of a minimum size far below (by a factor $\simeq$10) the minimum grain size expected from radiation pressure arguments. Such grains could not persist around the white dwarf due to its luminosity, so must be continually be created, or delivered, by some mechanism absent of a remnant planetesimal belt. The size distribution of Solar system cometary dust grains suggests a population of primarily sub-micron dust grains with a narrow size distribution \citep{2001Mason}. This would result in a narrow, steeply declining SED as has been observed here. A cometary origin for the dust grains in the system is therefore consistent with the continuum observations. 

However, the cometary cloud scenario might be expected to produce substantial amounts of volatiles such as water or carbon monoxide from volatile out-gassing from the planetesimals \citep[as seen in several debris discs;][]{2016Greaves,2017Kral,2020Marino, 2022Rebollido}.  The water ice line (150~K) lies at 28~au. This is passingly consistent with the inner edge of the dust emission region (35~au) as inferred from the \textit{Spitzer}/IRS spectrum. The CO ice line (30~K) lies at 740~au from WD~2226-210 ($T_{\rm eff} = 110\,000$~K, $L_{\rm WD} = 67~L_{\odot}$), well beyond the proposed location of the dust emission. Furthermore the dissociation timescale for CO in this environment would be very short, such that the amount of CO which could be produced by cometary outgassing, given the observed dust mass of 0.13~$M_{\oplus}$ \citep[and assuming a volatile composition similar to Solar system comets of 0.4-30~\%,][]{2011MummaCharnley}, would be rapidly photoionized and dissipated before it could build up into detectable levels around the white dwarf. 

%The non-detection of silicate emission features from small, warm dust grains at 10 and 20~$\mu$m in the \textit{Spitzer}/IRS mid-infrared spectrum must also be addressed by the cometary model. -- \jpm{I think the dust grains are too large and too cool to show a strong silicate feature based on expectations from debris discs (silicate features only observed in bright discs with dust at several 100 K, much hotter than seen here), so we don't need to go into this point, the lack of features in the IRS spectrum is, in fact, exactly what we'd expect.}

Remnant planetesimals from the main sequence system on more eccentric orbits will undergo a collisional cascade; the timescale for this evolution is dependent on the initial mass in planetesimals, the radial location of the planetesimal belt, and their distribution of eccentricities and inclinations. We can estimate the sublimation timescale for a planetesimal being irradiated by the white dwarf. Assuming a typical comet has a radius of 100~km, a perihelion distance of 30~au and an eccentricity of 0.5 we can calculate the sublimation timescale for a 100~km comet, typical for Solar system Edgeworth-Kuiper belt bodies, to be $<$~1,000~yrs \citep[see eqn. 11,][]{2015Stone}. This suggests that the observed dust emission is not the product of in-situ destruction of remnant planetesimals.

Alternatively, the observed dust emission is the result of cometary bodies on eccentric orbits that are returning to the inner parts of the system after being kicked onto high eccentricity orbits during the post main sequence evolution. A timescale of 10$^{5}$~yrs between post main sequence evolution and the current state of the Helix Nebula is comparable to the period of comets originating in the Solar system's Oort cloud and might point to the potential origin of this material. Given the observed dust mass of 0.13~$M_{\oplus}$ we can equate this to $\simeq~500\times10^{6}$ bodies with a mass equivalent to that of the Hale-Bopp comet, likely representative of an Oort cloud comet or mid-sized Kuiper belt object. Considering that the vast bulk of cometary bodies is icy matter rather than dust alone, this may be considered a conservative estimate of the number required. Further assuming that the dust we currently observe has been deposited over the past $10^{5}~$yrs without any loss due to radiation pressure, etc. therefore requires 5,000 Hale-Bopp equivalent comets per year being completely disrupted around the Helix Nebula white dwarf. Again, a conservative assumption as many cometary bodies will not be completely broken up. The observed dust must be deposited near-instantaneously and not built up over the duration by comets on relatively long periods as blow-out will remove dust which is generated mostly at periastron. If the cometary orbits are too short, thereby generating dust more frequently, then the comets won't survive the full $10^{5}~$yrs to produce the excess visible today. The disruption of a massive Kuiper belt during the post-main sequence evolution of the system, driving planetesimals to high orbital eccentricities, which are now still entering for only the first time, matches all the observable properties of the system. This leads to the preferred explanation of that excess emission originating from a disrupted planetary system.

 {If these are indeed cometary bodies returning to the inner parts of the Helix Nebula system, detecting the volatiles released by those evaporating comets would be one avenue to test the scenario we have proposed here \citep[e.g.][]{2020Rebollido}. However, this may be challenging against the bright and complex structure of the nebula. JWST MRS spectra of the white dwarf and its environment may yield evidence of dust features (e.g. crystalline or amorphous silicate), but there was no evidence for any spectral features in the relatively high signal-to-noise IRS spectrum. However, the higher angular resolution of JWST may help disentangle the circum-white dwarf environment from the nebula. Additionally, ELT/METIS N-band imaging of the circum-white dwarf environment could yield a detection of its emission, as the dust is deposited in a fairly well defined region around the white dwarf.}

\section{Conclusions}
\label{sec:conc}

We have modelled the emission from vicinity of the Helix Nebula's white dwarf (WD 2226-210) at infrared to millimetre wavelengths. We separate the emission into two physical components associated with the white dwarf for the modelling, based on structure seen in \textit{Herschel}/PACS observations at far-infrared wavelengths. Upper limits at far-infrared and millimetre wavelengths obtained respectively with SOFIA/HAWC+ and ALMA provide constraints on the spatial and emission properties of the dust associated with these physical components. 

The observed dust emission is inconsistent with either a remnant of the post-AGB envelope or the in-situ collisional destruction, or sublimation, of remnant planetesimals surviving from the main sequence system. The steep rise at mid-infrared wavelengths and absence of detectable sub-millimetre emission create a spectral shape of the excess emission inconsistent with the grain sizes expected from dust produced by planetesimals in a collisional equilibrium as would be expected in a classical debris disc scenario. Instead, we conclude that small dust grains must be present and that they need to be replenished at high rates due to the intense radiation field of the central white dwarf. 

We propose a scenario where the dust is released by heavy cometary activity around the white dwarf, demonstrating that the longevity of cometary bodies against sublimation necessitates a continual injection of dust by several thousand comets per year. This leads us to conclude that the excess emission observed around the central white dwarf in the Helix Nebula originates from the disruption of a massive Kuiper belt analogue.  This would likely have happened when a putative planetary system was destabilised during the star's post-main sequence evolution.

\begin{acknowledgements}
 {We thank the referee for their thoughtful review. J.P.M. thanks Alexander Mustill for his insight regarding the evolution of planetesimal belts in post-main sequence binary systems.} 

This research has made use of the SIMBAD database, operated at CDS, Strasbourg, France. This research has made use of NASA's Astrophysics Data System.

J.P.M. and F.K. acknowledge support by the Ministry of Science and Technology of Taiwan under grant MOST107-2119-M-001-031-MY3, and Academia Sinica under grant AS-IA-106-M03. J.P.M. acknowledges support by the Ministry of Science and Technology of Taiwan under grant MOST109-2112-M-001-036-MY3.

This work was also partly supported by the Spanish program Unidad de Excelencia Mar\'ia de Maeztu CEX2020-001058-M, financed by MCIN/AEI/10.13039/501100011033.

E.V. acknowledges support from the ``On the rocks II project'' funded by the Spanish Ministerio de Ciencia, Innovaci\`on y Universidades under grant PGC2018-101950-B-I00.

D.K. acknowledges  the  support  of  the  Australian  Research Council (ARC)  Discovery  Early  Career  Research  Award (DECRA) grant (DE190100813) and the Australian Research Council Centre of Excellence for All Sky Astrophysics in 3 Dimensions (ASTRO 3D), through project number CE170100013.

The work was partially based on observations made with the NASA/DLR Stratospheric Observatory for Infrared Astronomy (SOFIA). SOFIA is jointly operated by the Universities Space Research Association, Inc. (USRA), under NASA contract NNA17BF53C, and the Deutsches SOFIA Institut (DSI) under DLR contract 50 OK 0901 to the University of Stuttgart.  Financial support for this work was provided by NASA through award SOF\_05-0054\_Ertel issued by USRA.

This paper makes use of the following ALMA data sets: ADS/JAO.ALMA\#2015.1.00762.S and \#2016.1.00608.S. ALMA is a partnership of ESO (representing its member states), NSF (USA) and NINS (Japan), together with NRC (Canada) and NSC and ASIAA (Taiwan) and KASI (Republic of Korea), in cooperation with the Republic of Chile. The Joint ALMA Observatory is operated by ESO, AUI/NRAO and NAOJ.

This work benefited from the FEARLESS collaboration (FatE and AfteRLife of Evolved Solar Systems, PI: S. Ertel). The authors particularly acknowledge helpful discussions with Siyi Xu, whose input guided the formative stages of the project, and in obtaining the data.

\textit{Facilities:} ALMA, \textit{Herschel}, SOFIA, \textit{Spitzer}

\textit{Software:} This paper has made use of the Python packages {\sc astropy} \citep{2013AstroPy,2018AstroPy}, {\sc NumPy} \citep{2020Harris}, {\sc matplotlib} \citep{2007Hunter}.

\end{acknowledgements}

\bibliography{helix_refs}{}
\bibliographystyle{aasjournal}

\end{document}